\documentclass{INTERSPEECH2023}

\interspeechcameraready

\title{Influence of Utterance and Speaker Characteristics on the Classification of Children with Cleft Lip and Palate}
\name{Ilja Baumann$^1$, Dominik Wagner$^1$, Franziska Braun$^1$, Sebastian P. Bayerl$^1$, Elmar Nöth$^2$,\\ Korbinian Riedhammer$^1$, Tobias Bocklet$^{1,3}$}
\address{
  $^1$Technische Hochschule Nürnberg Georg Simon Ohm, Germany\\
  $^2$Friedrich-Alexander-Universität Erlangen-Nürnberg, Pattern Recognition Lab, Germany \\
  $^3$Intel Labs, Germany}
\email{firstname.lastname@th-nuernberg.de}

\begin{document}

\maketitle
 
\begin{abstract}
Recent findings show that pre-trained wav2vec 2.0 models are reliable feature extractors for various speaker characteristics classification tasks.
We show that latent representations extracted at different layers of a pre-trained wav2vec 2.0 system can be used as features for binary classification to distinguish between children with Cleft Lip and Palate (CLP) and a healthy control group.
The results indicate that the distinction between CLP and healthy voices, especially with latent representations from the lower and middle encoder layers, reaches an accuracy of 100\%.
We test the classifier to find influencing factors for classification using unseen out-of-domain healthy and pathologic corpora with varying characteristics: age, spoken content, and acoustic conditions. 
Cross-pathology and cross-healthy tests reveal that the trained classifiers are unreliable if there is a mismatch between training and out-of-domain test data in, e.g., age, spoken content, or acoustic conditions. 
\end{abstract}
\noindent\textbf{Index Terms}: pathologic speech, cleft lip and palate, children's speech

\vspace{-1mm}
\section{Introduction}
\label{sec:intro}

Speech pathologies have been studied using a wide variety of approaches.
Latent features such as GMM-Supervectors \cite{campbell06}, i-vectors \cite{dehak11_ivec} and x-vectors \cite{snyder18xvector} have been found useful for the analysis of pathologies from speech samples. 
Bocklet et al. use GMM-Supervectors to assess various speech pathologies \cite{bocklet09_AVFA}, to evaluate the intelligibility of laryngeal cancer patients \cite{bocklet12-jvoice} and to rate Parkinson's patients \cite{bocklet13_interspeech}.
X-vectors \cite{moro20_xvec_pd} were used for distinction among patients with Parkinson's disease and a healthy control group. 
More recent research employed latent representations obtained from wav2vec 2.0 \cite{baevski20w2v2} (W2V2) models in vocal fatigue detection, speech emotion recognition, and dysfluency detection \cite{pepino21_interspeech,bayerl22ksof,BAYERL2023vocalfatigue}. 
Furthermore, latent W2V2 representations have been successfully used to distinguish between speakers and languages \cite{Fan2021ExploringW2}.

We utilize wav2vec 2.0 (W2V2) to extract latent speech representations and explore their suitability for detecting Cleft Lip and Palate (CLP) in children's voices using two different kinds of binary classifiers: large-margin classifiers and neural networks.
Finally, we perform cross-corpora tests using out-of-domain data and investigate which utterance and speaker characteristics are important for CLP detection.
We alternate features such as age, spoken content, and acoustic conditions in a controlled manner to estimate their influence on classification performance.
We build upon the concerns raised in \cite{botelho22_interspeech} and analyze the claim that models often learn to distinguish recording conditions and speaker demographics.
To the best of our knowledge, this is the first work to examine W2V2 embeddings regarding various impact factors for children’s voice distinction using cross-pathologic and cross-healthy corpora tests. 

\vspace{-2mm}
\section{Data}
\label{sec:data}
We use a children's Cleft Lip and Palate (CLP) pathological speech corpus to train binary classifiers on features extracted using a pre-trained W2V2 encoder. The training data was divided into a training and test set (random 80\% / 20\% split).

The Erlangen-CLP corpus \cite{bocklet_14_clp} is a speech corpus of children with Cleft Lip and Palate (CLP) and age-matched control speakers (CLP\_C) recorded using the PLAKSS-II \cite{fox2014psycholinguistische} test (Psycholinguistische Analyse Kindlicher Sprechstörungen).
PLAKSS is a semi-standardized test consisting of words with all German consonants, vowels, and consonant clusters. 
Speech therapists widely use it in German-speaking countries.
Details are listed in Tables \ref{tab:patho-corpora} and \ref{tab:healthy-corpora}.

We perform cross-corpora tests using pathological datasets and additional corpora consisting of healthy speech to analyze how the trained CLP models perform on unseen data. 

\vspace{-1mm}
\subsection{Pathological speech corpora and controls}
\label{ssec:patho}
In the following Section, the pathological speech corpora are briefly described.
Details of the corpora can be found in Tables \ref{tab:patho-corpora} and \ref{tab:healthy-corpora}.

\noindent\textbf{Stegen-CI (SCI):}
The SCI corpus consists of audio recordings from children with hearing impairments and partially with Cochlear Implants (CI).

\noindent\textbf{Neumann/Fox-Boyer (NFB):}
The NFB clinical PhonBank corpus \cite{fox2014psycholinguistische} contains audio recordings of children with speech and sound disorders gathered from PLAKSS. Recordings were performed in different locations in Germany's western and northern regions.

\noindent\textbf{Laryng41 (LAR):}
The tracheoesophageal (TE) substitute voice is a typical treatment to restore the ability to speak after laryngectomy, i.e., the removal of the entire larynx. 
The LAR \cite{haderlein07} corpus is a collection of tracheoesophageal speakers reading the German version of the ``The North Wind and the Sun'' (NWS) text passage \cite{kohler_1990}. NWS is a phonetically rich text widely used in speech therapy.

\noindent\textbf{Oral Squamous Cell Carcinoma (OSCC):}
OSCC and its treatment impair speech intelligibility by alteration of the vocal tract. 
Patients were recorded reading the NWS text passage. 

\noindent\textbf{Parkinson's Disease (PD):}
Parkinson's disease is a degenerative disorder of the central nervous system, mainly affecting the motor system. 
The full PD corpus \cite{bocklet13_interspeech} contains native German speakers diagnosed with PD and a healthy control group. 
We use one of the tasks contained: reading a phonetically rich text. 

\noindent\textbf{AgedVoices110 (AV):}
\label{ssec:healthy}
Since LAR and OSCC have no matching controls, we use a corpus called AV consisting of elderly native German speakers reading the NWS text passage.
The AV corpus is similar to the LAR and OSCC corpora. It includes readings from a group of people roughly in the same age range ($>60$ years), recorded at the same hospital using the same equipment and speaking the same dialect.

\begin{table}[th]
  \vspace{-1mm}
  \caption{Pathological speech corpora details.}
  \label{tab:patho-corpora}
  \centering
  \vspace{-2mm}
  \begin{tabular}{rlllll}
    \toprule
    \textbf{Corpus}  & \textbf{Age}             & \textbf{Speaker}    & \textbf{Hours} & \textbf{Content} \\
    \midrule
    CLP     & $8.7 \pm 13.3$    & 332           & 28.9  & PLAKSS \\
    \midrule
    SCI     & $13.4 \pm 6.4$    & 193           & 9.2   & PLAKSS \\
    NFB     & $5.2 \pm 4.4$     & 29            & 0.7   & PLAKSS \\
    LAR     & $62.0 \pm 7.7$    & 41            & 0.7   & NWS text \\
    OSCC    & $59.9 \pm 10.1$   & 71            & 1.0   & NWS text \\
    PD      & $66.6 \pm 9.0$    & 88            & 1.2   & Text \\
    \bottomrule
  \end{tabular}
\end{table}

\begin{table}[th]
    \vspace{-4mm}
    \caption{Control and healthy speech corpora details. Sentences$^1$ and Sentences$^2$ differ in content.}
  \label{tab:healthy-corpora}
  \centering
  \vspace{-2mm}
  \begin{tabular}{rlllll}
    \toprule
    \textbf{Corpus}  & \textbf{Age}             & \textbf{Speaker}    & \textbf{Hours} & \textbf{Content} \\
    \midrule
    CLP\_C     & $8.7 \pm 13.3$    & 598       & 39.5  & PLAKSS \\
    AV      & $75.7  \pm 9.6$   & 110       & 1.7   & NWS text \\
    PD\_C      & $58.1 \pm 14.2$   & 88        & 1.2   & Text \\
    \midrule
    RH      & $11.6 \pm 7.2$    & 177       & 1.9   & RHINO \\
    FB      & $5.1 \pm 4.1$     & 32        & 0.7   & PLAKSS \\
    PSZ     & $16.0 \pm 4.0$    & 10        & 1.0   & Sentences$^1$ \\ 
    TUDA    & -                 & 10        & 1.0   & Sentences$^2$ \\
    NWSR    & $66.6 \pm 9.0$    & 8         & 0.6   & NWS text \\
    MCK     & $4.5 \pm 1.5$     & 2         & 0.1   & NWS \& \\
            &                   &           &       & PLAKSS \\
    \bottomrule
  \end{tabular}
  \vspace{-4mm}
\end{table}

\subsection{Additional healthy corpora}
\label{ssec:additional}
The following Section lists the additional healthy corpora used for cross-tests.
Details can be found in Table \ref{tab:healthy-corpora}.

\noindent\textbf{RHINO (RH):}
The dataset consists of children who spoke the RHINO (Heidelberger Rhinophoniebogen) test, which was developed to assess nasality. 
The dataset consists of the identical children of the Erlangen-CLP control group and additional speakers. 
The utterances contain sustained vowels, consonants, and sentences with and without nasal consonants.

\noindent\textbf{Fox-Boyer (FB):}
The FB corpus \cite{fox2014psycholinguistische} contains audio recordings of typically developing children gathered from PLAKSS. Recordings were performed in different kindergartens in the northeast region of Germany or private practices in the western and northern regions of Germany.

\noindent\textbf{Phattsessionz100 (PSZ):}
The PhattSessionz project \cite{draxler05_interspeech} is a regionally balanced speech database of German adolescent speakers. 
We randomly selected 10 speakers from the ``reading phonetically rich sentences'' task for our experiments. 

\noindent\textbf{Tuda-De100 (TUDA):}
The TUDA data comprises 100 utterances from the Tuda distant speech corpus \cite{radeck_15_tuda}, aged 21 to 30. 
We randomly selected a subset of utterances from 10 speakers (5 female, 5 male) recorded using a Yamaha PSG-01S microphone. 
Each speaker reads 10 sentences from German Wikipedia. 


\noindent\textbf{NWS Reading (NWSR):}
NWSR is a small corpus comprising 8 native German speakers reading the NWS text passage multiple times throughout approximately one year. 
Six speakers were older than 50, and two were 12 and 23.

\noindent\textbf{Multi-content Kids (MCK):}
MCK consists of two children reading the NWS text passage and performing the PLAKSS test.

\section{Method}
\label{sec:method}

\subsection{Latent features}
\label{subsec:embeddings}
\vspace{-1mm}
We use a pre-trained (LibriSpeech) W2V2 base model fine-tuned for ASR to extract embeddings for each utterance of the individual datasets from each of the 12 transformer blocks over the entire utterance length.
The dimensionality of the resulting features is 768. The receptive field is 25ms with a stride of 20ms.

We first analyzed the W2V2 embeddings of the CLP dataset at different levels: aggregated by utterance, speaker, and layer using the mean. Based on available metadata, we further investigated whether grouping by in-common properties of the embeddings occurs by visualizing embeddings using t-SNE plots.
Examined metadata age, gender, recording location, accent, and spoken text.
Furthermore, based on previous works \cite{pepino21_interspeech,cormac-english-etal-2022-domain,baumann22_interspeech}, we assume that the lower encoder blocks contain low-level features; the higher the layer, the more content and phonetic information these contain.
For this reason, we aggregate three layers at a time, calculating the average over the lower layers (1-3), middle layers (4-6 and 7-9), and higher layers (10-12).

\begin{figure}[!htb]
    \begin{center}
	\includegraphics[width=0.88\columnwidth]{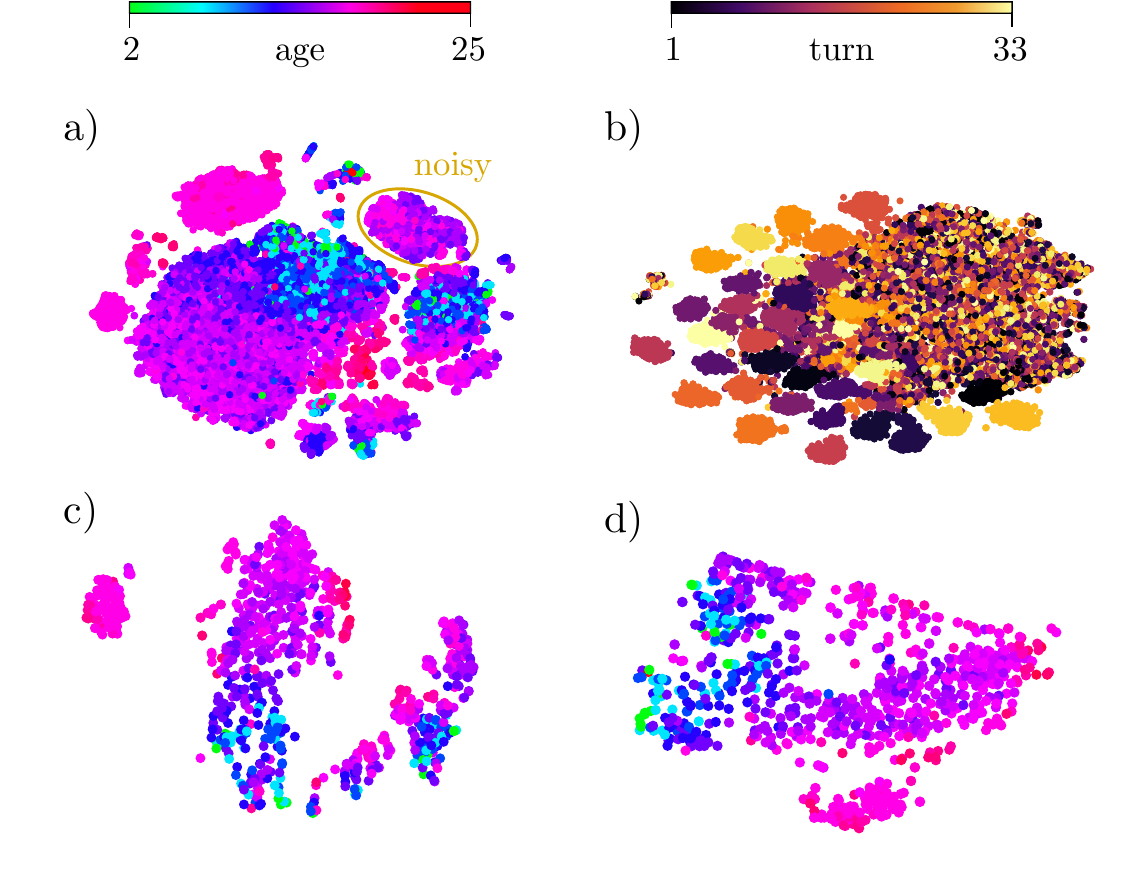}
	\vspace{-2mm}
    \caption{t-SNE projection ($perplexity=30$) of CLP features extracted and aggregated at layers 1-3 (a, c) and 7-9 (b, d) and utterance level (a, b) vs. speaker level (c, d). The top-left and bottom plots show the age distribution, and the top-right plot shows the PLAKSS test turn clusters.}
  \label{fig:utt-vs-spk}
  \end{center}
\end{figure}

Figure \ref{fig:utt-vs-spk} shows the embeddings at utterance level in a) and b) and at speaker level in c) and d). Further, a) and c) were extracted from lower layers 1-3 and b) and d) from the middle layers 7-9.\footnote{All plots can be explored at:  https://clpclf.github.io/clp-clf}
Utterance embeddings from lower layers in a) show coarse clusters by acoustic conditions (noisy environment, above-average age) and continuous mapping of age structures. Color mappings correspond to age groups per year.
Utterance embeddings from middle layers in b) show for the most part distinct clusters by the spoken content, where each color represents the same three spoken words of the PLAKSS test.
A single turn in the PLAKSS test of the CLP corpus corresponds to three spoken words. Thus, for example, a turn contains the German words: \textit{Mond, Eimer, Baum}.
These findings are consistent with previous works, that more content information is found in the higher layers.
Utterance embeddings averaged to speaker embeddings in c) and d) show similar main groups as in a) by acoustic conditions and age structures, even in higher layer embeddings.
For further analysis, we aggregated the datasets by the speaker using the mean of all utterance embeddings of a speaker to obtain more robust features.

\vspace{-9mm}

\subsection{Classifier}
\label{subsec:classifier}
We evaluate two different kinds of models; large-margin classifiers and neural networks. 
We choose Support Vector Machine (SVMs) and a feedforward neural network with fully connected layers (FFN) for the binary classification. 
The optimal hyperparameters for each estimator are determined in a 5-fold cross-validation on the training set using the grid search method.

SVMs are trained using radial basis function (RBF) kernels. 
The kernel parameter $\gamma$ is selected from $\gamma \in \lbrace 10^{-k} \,|\, k = 5, \ldots , 1\rbrace \subset \mathbb{R}_{>0}$, and the penalty parameter of the error term $C$ is selected from $C \in \lbrace 5, 10, 20, 50 \rbrace \subset \mathbb{N}_{>0}$.


The FFN employs the Adam \cite{kingma14_adam} optimizer with exponential decay rates of $\beta_1 = 0.9$, $\beta_2 = 0.999$ and a $\mathcal{L}_2$ regularization term of $10^{-4}$. 
The FFN learning rate $\alpha$ is chosen from $\alpha \in \lbrace 10^{-k} \,|\, k = 1, \ldots , 4\rbrace \subset \mathbb{R}_{>0}$. 
The activation function is either Tanh or ReLU, the number of hidden layers is 2 or 3, and the number of hidden units varies between 32 and 128. 

\vspace{-1mm}
\subsection{Decision boundary plot}
\label{subsec:decisionboundary}
We use a modified version of high-dimensional decision boundary plot\footnote{https://github.com/tmadl/highdimensional-decision-boundary-plot} to analyze the classifier results and plot an approximate projection of the classifier decision boundaries.
For this purpose, samples from both classes are used to find the regions with the maximum uncertainty (probability=0.5) and locate keypoints, along connecting lines between the embeddings. 
Subsequently, along lines connecting the keypoints, maximum uncertainty regions are located; for the case of concave regions, uncertainty regions are located along hypersphere surfaces. 
Finally, the points are projected into two-dimensional space using t-SNE.

\vspace{-2mm}
\section{Experiments}
\label{sec:experiments}

\subsection{CLP voice classification}
\label{subsec:binary-classification}

We first train binary classifiers (SVM and FFN) to distinguish between CLP and matched control group. Inputs of the classifiers are the 12 W2V2 transformer block outputs extracted from audio and aggregated as described in \ref{subsec:embeddings}.

Table \ref{tab:clf-results} shows lower-layer, middle-layer, and high-layer results of the SVC and FFN classifier. 
The results indicate that the lower-layer and middle-layer embeddings are best suited for classifying the pathologic voices, resulting in accuracy scores from 94.6\% to 100\%. The scores are statistically significant at a significance level of $\alpha=0.05$.
The CLP classifiers yield similar results on lower-layer and middle-layer features, resulting in an approximate 5\% relative accuracy drop on high-layer features.
The results indicate that important pathological voice information can be represented by W2V2 embeddings, especially in the lower and middle layers. 
This is, for example, consistent with the results from \cite{Fan2021ExploringW2}, where the authors found that the lower-layer representations are best suited for speaker- and language-based distinction tasks.
Since the classifiers yield similar results, we will only consider the SVM results without having to average across classifier kinds.

\begin{table}[th]
    \caption{Classification accuracy results of the binary classifier for CLP, trained on features from lower-, middle-, and high-layers.}
  \label{tab:clf-results}
  \vspace{-2mm}
  \centering
  \begin{tabular}{rcccc}
    \toprule
    \textbf{Classifier} & \textbf{1-3}  & \textbf{4-6}  & \textbf{7-9}  & \textbf{10-12} \\
    \midrule
    SVC                 & 100           & 100           & 99.5          & 97.3 \\
    FFN                 & 100           & 99.5          & 99.5          & 94.6 \\
    \bottomrule
  \end{tabular}
  \vspace{-4mm}
\end{table}

\subsection{Cross-pathology tests}
\label{subsec:cross-patho-tests}
Since the binary classifiers yield near-perfect results on the CLP pathologic voice classification task, we investigate what was learned and how models trained on one pathology perform on out-of-domain pathological and healthy data.
We use the previously trained classifiers of the CLP pathology and test each of the five pathological datasets (SCI, NFB, LAR, OSCC, and PD) to see how these handle unseen pathologic speech that is not CLP. 
The evaluations are listed in Table \ref{tab:cross-patho-results}.
We report the results in percent classified as CLP pathologic voice.

CLP models classify the four pathologic corpora NFB, LAR, OSCC, and PD as CLP with an accuracy of 100\% trained on the lower-layer embeddings. Middle-layer embeddings (4-6) result in uncertainty for LAR (56.8\%) and NFB (65.5\%), whereas OSCC and PD are classified as 71.2\% and 85.2\% as CLP.
SCI speakers are predicted as CLP on high-layer features while on all other embeddings as healthy speakers.
Classifiers trained on high-layer embeddings predict all pathologic corpora as CLP.

It is unclear why the CLP models classify all other pathologies except CI as CLP on lower-layer and middle-layer features. 
This could be due to the significant age difference or acoustic differences between the corpora, which leads to the classification of unseen data into the CLP class, especially on lower-layer features. 
Moreover, the spoken content also differs between the corpora, which would explain the variability in high-layer features. 
Whereas in CLP, single words are uttered, in LAR and OSCC, the NWS text is read, and in PD, a completely different text is read.

\begin{table}[th]
  \caption{Cross-corpora SVM classification results for pathological data. }
  \label{tab:cross-patho-results}
  \vspace{-1mm}
  \centering
  \begin{tabular}{rcccc}
    \toprule
    \textbf{Corpus} & \textbf{1-3}  & \textbf{4-6}  & \textbf{7-9}  & \textbf{10-12} \\
    \midrule
    NFB             & 100           & 65.5          & 34.5          & 100 \\
    SCI             & 0             & 7.9           & 1.7           & 100 \\
    LAR             & 100           & 56.8          & 0             & 100 \\ 
    OSCC            & 100           & 71.2          & 34.6          & 100 \\
    PD              & 100           & 85.2          & 34.1          & 100 \\
    \bottomrule
  \end{tabular}
  \vspace{-4mm}
\end{table}

\subsection{Cross-healthy tests}
\label{subsec:cross-healthy-tests}
Similar to cross-pathology tests, we conduct tests on unseen out-of-domain data. Additional to the training controls data, we evaluate the CLP pathology classifiers on the corpora described in Section \ref{ssec:additional}.
The results are listed in Table \ref{tab:cross-healthy-results}.
As in Section \ref{subsec:cross-patho-tests}, we report the results as the percentage classified as CLP.



No consistent result emerges by looking at the predictions of the CLP classifiers, except for high-layer features where all corpora are entirely classified as CLP.
The results of the classifiers trained on low- and middle-layer features do not indicate bias according to age structure. Young children in FB: 100\% CLP; in contrast, the elderly speakers in AV: 20-28.3\% CLP. 
A similar result is obtained for spoken content. For example, the lower-layer feature classifier yields a large discrepancy for PD (phonetically rich text, 100\% as CLP) and NWSR (NWS text, 18.8\% as CLP).
It is also remarkable that despite the same spoken content (PLAKSS) and a shared age range with the training data, the entire FB corpus is classified as CLP regardless of the used training features layers of the classifier.


\begin{table}[t]
  \caption{Cross-healthy SVM classification accuracy results for controls and out-of-domain healthy data.}
  \label{tab:cross-healthy-results}
  \vspace{-2mm}
  \centering
  \begin{tabular}{rccccc}
    \toprule
    \textbf{Corpus}  & \textbf{1-3}    & \textbf{4-6}   & \textbf{7-9}   & \textbf{10-12} \\
    \midrule
    PD      & 100   & 69.3  & 18.2  & 100 \\
    AV      & 28.2  & 20.0  & 1.8   & 100 \\
    NWSR    & 18.8  & 38.5  & 23.9  & 100 \\
    TUDA    & 100   & 92.2  & 23.5  & 100 \\
    PSZ     & 30.0  & 40.0  & 30.0  & 100 \\
    FB      & 100   & 100   & 100   & 100 \\
    MCK     & 100   & 0     & 0     & 100 \\
    RH      & 5.6   & 88.7  & 65.5  & 100 \\
    \bottomrule
  \end{tabular}
  \vspace{-5mm}
\end{table}

\subsection{Analysis}
To better understand how classification results are produced on out-of-domain data, we examine the most prominent differences between the corpora: acoustic conditions, age structure, and spoken content. 

\vspace{-1mm}
\subsubsection{Acoustic conditions}
Considering only CLP\_C from Figure \ref{fig:tsne-sv}, three clusters are evident. The small bottom cluster refers to older children ($\mu = 12.1  \pm 4.7$ years old, overall corpus $\mu = 7.8$) and is recorded in a different room. While the right top small cluster consists of speaker recordings with noticeable noise in the background (SNR $= 7$) compared to the remaining dataset recordings (SNR $= 22$).
Especially embeddings from the lower layers show the influence of the acoustic conditions.
The FB corpus was recorded using an Olympus WS-650S digital audio recorder or a later version, probably placed on the table in front of the children. In contrast, the whole CLP and RH corpora were recorded using a standard headset microphone (Plantronics Audio .655).

\vspace{-1mm}
\subsubsection{Spoken content}
The spoken content across the corpora is divided into PLAKSS words (CLP, SCI, FB), read sentences (TUDA, PSZ), read phonetically rich text (PD), NWS text (NWSR, MCK, AV, LAR, OSCC) and single vowels and consonants combined with read sentences (RH). Figure \ref{fig:tsne-sv} shows all corpora used with the extracted lower-layer (1-3) average embeddings.
The lower area of the Figure shows all PLAKSS corpora, including CLP, SCI, FB, and additionally RH. At the top area, a group of corpora OSCC, AV with the read NWS passage, and PD with the read text is formed ($>$50 years). These clusters could also arise from age differences, but tests with MCK show that for lower-layer embeddings, both children are assigned to the child embeddings of the other corpora (lower area), regardless of the spoken content. Regarding middle-layer embeddings, both children are located at the NWS corpora (OSCC, LAR) using NWS embeddings and remain at the bottom area using PLAKSS features.
Our analysis of all aggregated layers shows that spoken content is encoded in the latent features, especially in middle-layers.


\vspace{-1mm}
\subsubsection{Age structures}
The age structures vary substantially across the used datasets. The youngest children are included in FB and MCK, CLP, CI, and PSZ for the children/adolescent corpora. TUDA follows with 21 to 30 years old speakers. The oldest speakers are included in NWSR, LAR, OSCC, and AV, with $>$50 years old.
Acoustic near representations are recognizable especially based on the lower-layer features. Children of any spoken content are grouped. This observation shifts in the middle- and high-layer features, where the spoken content seems to have a stronger effect.

\vspace{-1mm}
\subsubsection{Decision boundary}
\begin{figure}[!htb]
	\includegraphics[width=0.98\columnwidth]{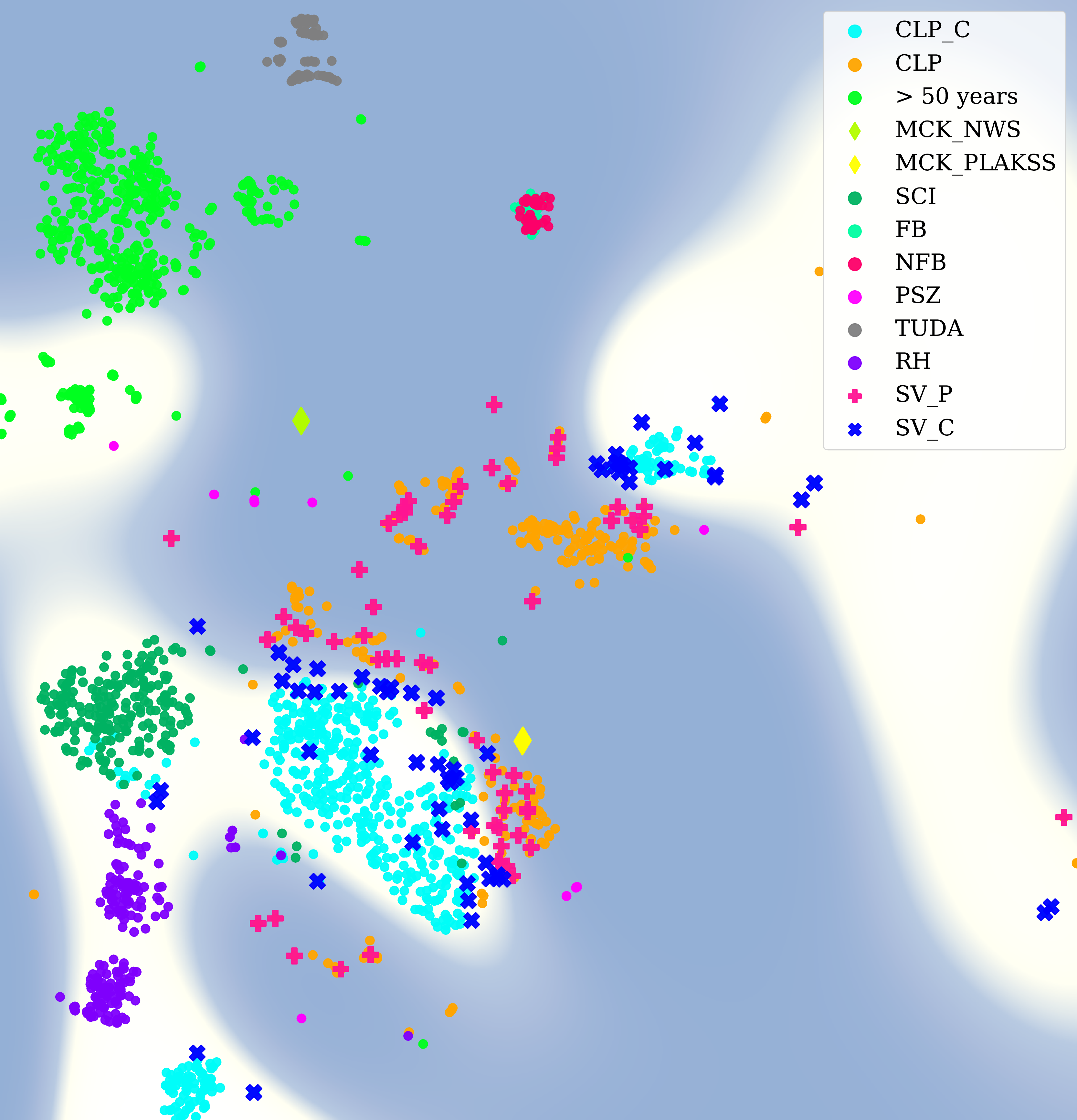}
    \caption{2-dimensional t-SNE projection ($perplexity=90$) of the mean latent features extracted at W2V2 lower-layers (1-3) encoder for control datasets. (\_C = controls, \_P = pathologic). The background approximates the SVM decision boundary; blue represents the CLP class, white the healthy control. }
  \label{fig:tsne-sv}
\end{figure}

To gain additional insights into how the results from Tables \ref{tab:cross-patho-results} and \ref{tab:cross-healthy-results} emerge, we attempt to visualize the decision boundary according to the procedure described in Section \ref{subsec:decisionboundary}. 
Figure \ref{fig:tsne-sv} shows an approximation of the SVM decision boundary as a contour plot in the background; blue represents the CLP class, and white is the healthy control group.
We further extract the support vectors from the SVM classifiers. 
Embeddings, decision boundary points, and support vectors were simultaneously projected into a 2-dimensional space for visualization. Figure \ref{fig:tsne-sv} shows the support vectors of the pathological class with an "+" marker and the healthy class with an "x" marker. 

The decision boundary, especially in combination with the support vectors, suggests that there is not always a clear boundary to features not seen in the training. 
Corpora with unseen age structures, spoken content, and acoustic features are classified ambiguously.
Therefore, it is important to include as many of the aforementioned stimuli and conditions as possible in the training data to cover as many uncertainties as possible and to create robust classifiers that generalize to out-of-domain data.

\vspace{-5mm}
\section{Conclusions}
\label{sec:conclusion}
We have shown that latent representations extracted using a pre-trained W2V2 model include characteristics necessary for children's speech distinction with cleft lip and palate.
Especially the features extracted from the low and middle layers exhibit the strongest distinction required for CLP speech detection.
The best classifier accuracy trained on low-layer and middle-layer features is 100\%.
However, based on the cross-corpora tests, selecting diverse speech stimuli and conditions is important for robust classification, especially in binary classification. 
The influence also differs depending on the extracted layer of the embeddings.
The trained pathology classifiers provide ambiguous results on out-of-domain data because only one spoken content (PLAKSS), the same recording equipment, and only children were part of the training data.
Our findings support the concerns raised in \cite{botelho22_interspeech}. However, our multi-class extension \cite{multi_patho_clf22} of this work  shows that classifiers trained on different pathologies and characteristics in a multi-class scenario provide robust results and have a regularizing effect.

\bibliographystyle{IEEEtran}
\bibliography{mybib}

\end{document}